\documentclass[10pt,aps,pre,twocolumn,superscriptaddress]{revtex4-2}
\usepackage{amsmath, amssymb, graphicx}
\usepackage{xcolor}
\usepackage[euler]{textgreek}
\usepackage{placeins}
\usepackage[hidelinks]{hyperref}
\usepackage{cleveref}

\crefname{figure}{figure}{figures}
\Crefname{Figure}{Figure}{Figures}
\crefname{equation}{equation}{equations}
\Crefname{Equation}{Equation}{Equations}

\begin{document}

\title{Performance of Nanoring-based Transparent Conductors: \texorpdfstring{\\}{}a Computational Investigation}

\author{Gijs Vanoppen}
\affiliation{UHasselt-Hasselt University, Faculty of Sciences, Theory Lab, Agoralaan, 3590 Diepenbeek, Belgium}
\author{Jef Hooyberghs}
\affiliation{UHasselt-Hasselt University, Faculty of Sciences, Theory Lab, Agoralaan, 3590 Diepenbeek, Belgium}
\affiliation{UHasselt-Hasselt University, Faculty of Sciences, Data Science Institute, Agoralaan, 3590 Diepenbeek, Belgium}
\author{Wim Deferme}
\affiliation{UHasselt, Institute for Materials Research, Martelarenlaan 42, 3500 Hasselt, Belgium}
\affiliation{IMEC vzw, Division IMOMEC,  Wetenschapspark 1, 3590 Diepenbeek, Belgium}
\author{Bart Cleuren}
\email[]{bart.cleuren@uhasselt.be}
\affiliation{UHasselt-Hasselt University, Faculty of Sciences, Theory Lab, Agoralaan, 3590 Diepenbeek, Belgium}

\date{\today}

\begin{abstract}
Metallic nanoring networks can serve as promising flexible transparent electrodes.
These materials are crucial components in a wide range of applications, including solar cells, touchscreens and displays. In this work, a computational investigation considers in detail (i) the electrical conductance and optical performance of nanoring networks and (ii) the breakdown of these networks due to electrical damage. The electrical resistance of both the nanorings and the contacts between the rings (junctions) is taken into account. In part (i), the effects of 5 parameters on the electrical sheet resistance and optical transparency are presented. It is shown that several parameter combinations achieve better performance in comparison to indium tin oxide, currently the most widely used transparent electrode. In part (ii), due to electrical damage, the nanoring systems display the formation of a crack, running parallel to the vertical terminals, where a voltage difference is applied. The network degradation is measured by its sheet resistance, and a universal effect is observed: networks with varying filling factors exhibit the same degradation profile.
\\$ $
\end{abstract}

\maketitle

\section{Introduction}
Flexible transparent electrodes (FTEs) are essential elements in emerging optoelectronic technologies, including wearable devices, organic photovoltaics, supercapacitors, liquid crystal displays, transparent heaters and next-generation touch interfaces \cite{wang_recent_2025, kim_revisiting_2018, maji_tailoring_2025, wan_enhanced_2024}.  These materials should combine optical transparency, electrical conductance and mechanical flexibility while maintaining stability in their operating environments. Unfortunately, current industry-standard electrodes do not satisfy all of these requirements. The prime example is indium tin oxide (ITO). ITO is the most widely used transparent electrode as it is both highly transparent and conductive \cite{minami_transparent_2005}. However, due to the brittle nature of this material, it is not a suitable choice for flexible applications \cite{yun_fabrication_2013, witte_strain-dependent_2000}. Additionally, due to high production costs and the scarcity of indium, finding alternatives has become a research hotspot. 

Both metallic nanowire and nanoring networks have proven to be primary candidates for replacing ITO films in devices such as organic light-emitting diodes and solar cells \cite{azani_benefits_2020, huang_metal_2022, verboven_ultrasonic_2022}. This is due to their ability to leverage the outstanding conductivity of metals like silver and copper, while still displaying high transparency.
Additionally, these networks can operate in environments that require flexibility \cite{kim_roll--roll_2016}. Specifically, nanoring networks appear to be highly enticing as FTEs due to their unique geometry. For instance, every part of the percolating cluster (to be explained further in this section) will carry some non-zero current, meaning there are no dead ends in the system, as is the case in nanowire networks \cite{han_computational_2018}. One can expect that this leads to a more uniform current distribution, reducing the number of hotspots in the network and therefore benefiting the network stability. Additionally, owing to their geometry, overlapping rings always have two points of contact. Hence, it has been speculated that more wire-wire contacts might be present as compared to the nanowire case, leading to a reduction in sheet resistance \cite{azani_transparent_2019}. Indeed, using a solvothermal method, Azani et al. \cite{azani_silver_2018} created nanoring networks that exhibited higher transparency at an equal sheet resistance compared to their nanowire counterparts.

When studying connectivity by randomly deposited objects on a plane, such as nanowire and nanoring networks, one enters the field of percolation theory. In 2D continuum percolation theory, when depositing $N$ objects in a square system of size $L\times L$, the number density $n$ is defined as
\begin{equation}
    n = \frac{N}{L^2}.
\end{equation}
The number density can be used to define the percolative filling factor $\eta$ \cite{mertens_continuum_2012}, which is simply the result of multiplying the number density $n$ by the area $a$ of the object,
\begin{equation}\label{eq:filling_factor}
    \eta = na.
\end{equation}
For rings of radius $r$, $a=\pi r^2$. The filling factor is a very useful quantity, as it is dimensionless and indicates the density of the network, regardless of the system size $L\times L$.

In a nanoring system, multiple sets of interconnected rings, called clusters, can be formed. A system is said to be percolating when at least one spanning cluster is present, that is a cluster containing a conductive pathway between the two connecting electrodes (i.e. between the two busbars in \cref{fig:gen_perc_cur}(b)). In the limit $L\rightarrow\infty$, a critical value $\eta_c$ (called the percolation threshold) of the filling factor is identified, where the probability of finding a spanning cluster is zero when $\eta < \eta_c$ and unity when $\eta > \eta_c$. At finite sheet sizes, this Heaviside function relaxes. However, to ensure that the network conducts well, relevant industrial applications will always have a filling factor larger than $\eta_c$. For continuum ring percolation, the percolation threshold is \mbox{$\eta_c$ = 1.12808737(6)} \cite{mertens_continuum_2012}. 
For continuum percolation of 1D sticks with length $l$ (which can be used to model straight nanowires), $a$ is defined as $l^2$ and the percolation threshold is \mbox{$\eta_c$ = 5.63726(2) \cite{mertens_continuum_2012}}.

When nanowire or nanoring networks are in use, an electric current will flow through the conducting material in the system. As these conducting wires have a small cross-section, the current density can become very large, causing the wires to break down as a result of, for example, electromigration or the Joule effect \cite{lagrange_understanding_2016, yagi_quantifying_2019, resende_time_2022}. This destruction on a local scale results in the increase of the global sheet resistance. Beyond a critical level of damage, the network can even stop percolating, meaning the electrode no longer has the ability to conduct electrical charges. Understanding and preventing this electrical breakdown is paramount for the fabrication of stable networks for their implementation in durable devices.

Previous computational studies of nanoring network performance do not account for ring thickness, transparency or junction resistance, either individually or a combination thereof. Additionally, to the best of the authors' knowledge, no previous attempts have been made to investigate the electrical breakdown of nanoring networks. This contribution aims to address these two aspects. Specifically, we present a computational investigation that inspects (i) the performance of nanoring networks, focusing on the balance between transparency and sheet resistance, and (ii) the breakdown of these networks due to electrical damage. 

\section{Methods: modeling and simulation}\label{s:methods}
In this work, a homemade \verb|C++| code is developed to simulate conductive nanoring networks. 
These simulations allow for the calculation of both the sheet resistance and transparency, as well as to investigate the effect of electrical damage to the network. This section gives an overview of important details about the simulation. Subsection \ref{ss:Geometrical_considerations} describes the generation of the nanoring networks and their clustering strategy. Then, subsection \ref{ss:R-T_performance} illustrates how a generated network is used to calculate its sheet resistance $R_\square$ and transparency $T$. Finally, subsection \ref{ss:electrical_damage} shows how the simulation handles the breakdown of the networks due to electrical damage.

\subsection{Geometrical considerations}\label{ss:Geometrical_considerations}
\begin{figure*}
    \centering
    \includegraphics[width=0.9\linewidth]{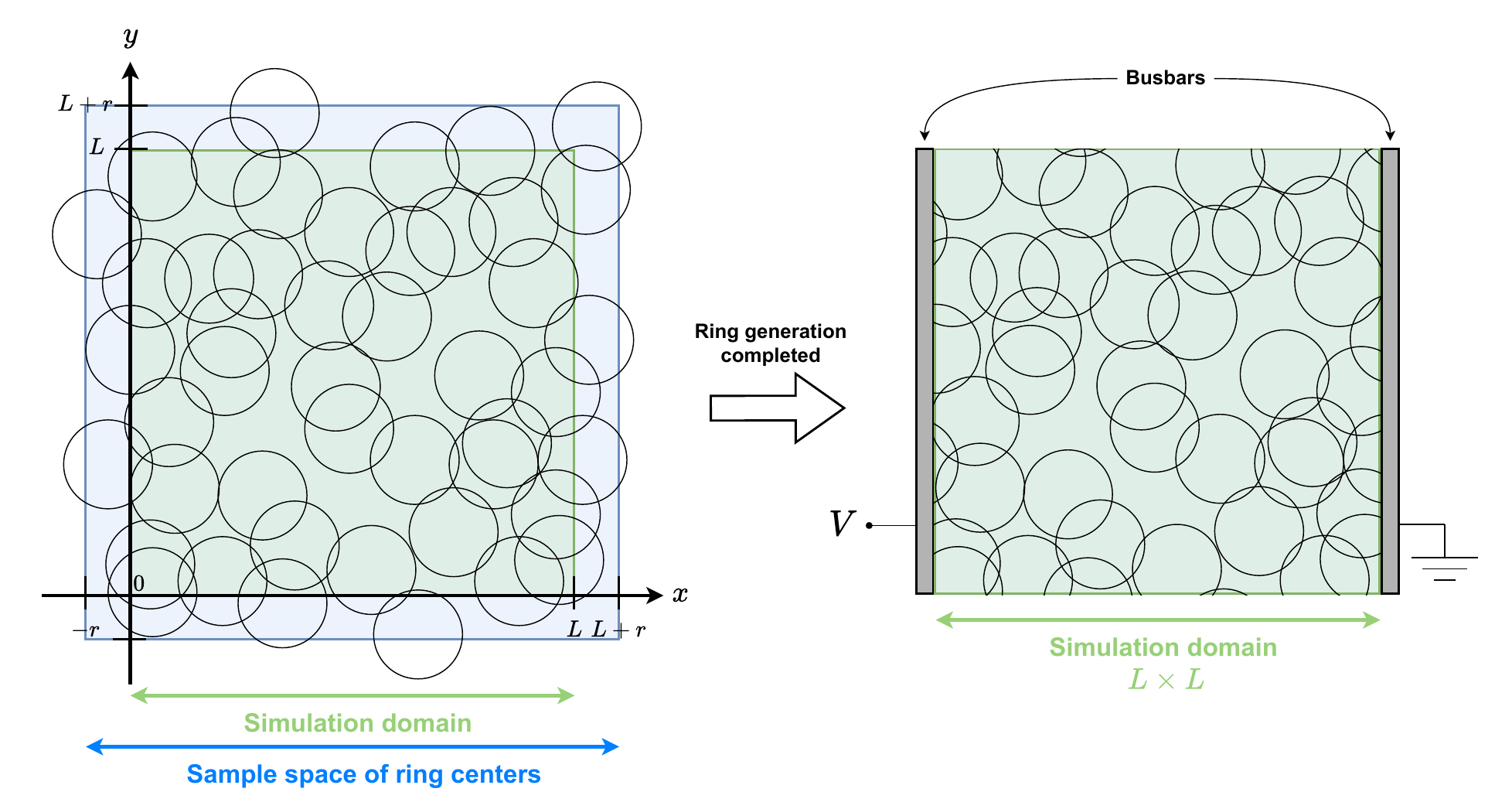}
    \caption{During nanoring generation, the coordinates of the centers are sampled from the blue+green space. After generation, only the green domain is considered for the rest of the simulation. Superconducting busbars are constructed to the vertical boundaries, and a potential difference $V$ is applied. }
    \label{fig:simulation_domain}
\end{figure*}

Nanorings with radius $r$ are deposited on a square simulation domain of size $L\times L$. The coordinates of the nanoring centers are identically and independently drawn from a uniform distribution in the domain  \mbox{$]-r, L + r[$}. After $N$ rings are deposited, everything in the simulation outside of the domain $[0,L]$ is removed, as shown in \cref{fig:simulation_domain}.

Using this method, the rings are allowed to cross the boundary of the sheet.
This approach differs from the usual method of applying periodic boundary conditions (PBC), as seen in e.g. \cite{tarasevich_electrical_2021, tarasevich_transparent_2023, azani_transparent_2019}. 
Using these PBC can introduce spatial correlations in the distribution of the ring center coordinates.
After the generation of nanorings is completed, a busbar with negligible resistance is connected to both vertical boundaries, and a potential difference $V$ is applied. 
This will be used to measure the conductive properties of the system, as will be explained in \cref{ss:R-T_performance}.

\begin{figure*}
    \centering
    \includegraphics[width=0.8\linewidth]{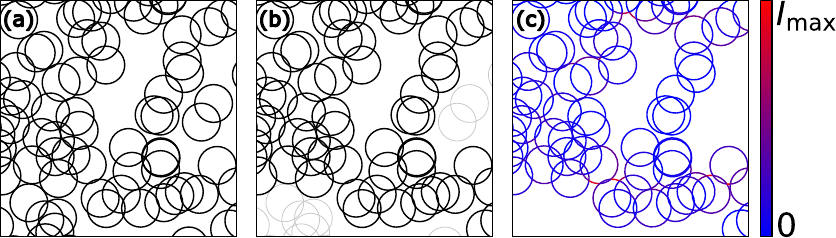}
    \caption{(a) A set of generated nanorings, created by the procedure in \cref{ss:Geometrical_considerations}.
    (b) After application of the Union-Find algorithm on the sample in (a), the spanning cluster (black rings) has been identified.
    The other clusters are grayed out. (c) The distribution of electrical current throughout the percolating cluster.}
    \label{fig:gen_perc_cur}
\end{figure*}

After the desired number of rings is generated, a Union-Find algorithm (adapted for continuous systems) is used to find a spanning cluster. A cluster is spanning when it is connected to both busbars. A result of this cluster finding algorithm is illustrated in \cref{fig:gen_perc_cur}(a) and (b).

Note that, although unlikely (especially at high ring densities), multiple spanning clusters can coexist in one sample. In that case, the subsequent methods are applied to all spanning clusters in the sample.

Also note that when calculating whether two rings intersect or not, it is assumed that the nanorings are objects with zero width.
Given that nanorings are high-aspect-ratio objects, this choice is well justified.
Avoiding this assumption would introduce substantial complexity in calculating the Equivalent Electrical Circuit in \cref{ss:R-T_performance}.

To improve the computational performance of calculating the presence of overlapping rings, spatial partitioning with a uniform grid is utilized: the simulation domain is divided into square grid cells with a size equal to the diameter of the rings.
Each grid cell keeps track of which rings intersect with the cell.
Then, when calculating which rings in the system overlap with a certain other ring, only the rings in nearby grid cells need to be checked instead of every ring in the whole system.
This makes the clustering algorithm many times more efficient.

\subsection{Calculating conductive and optical performance} \label{ss:R-T_performance}
\begin{figure*}
    \centering
    \includegraphics[width=0.7\linewidth]{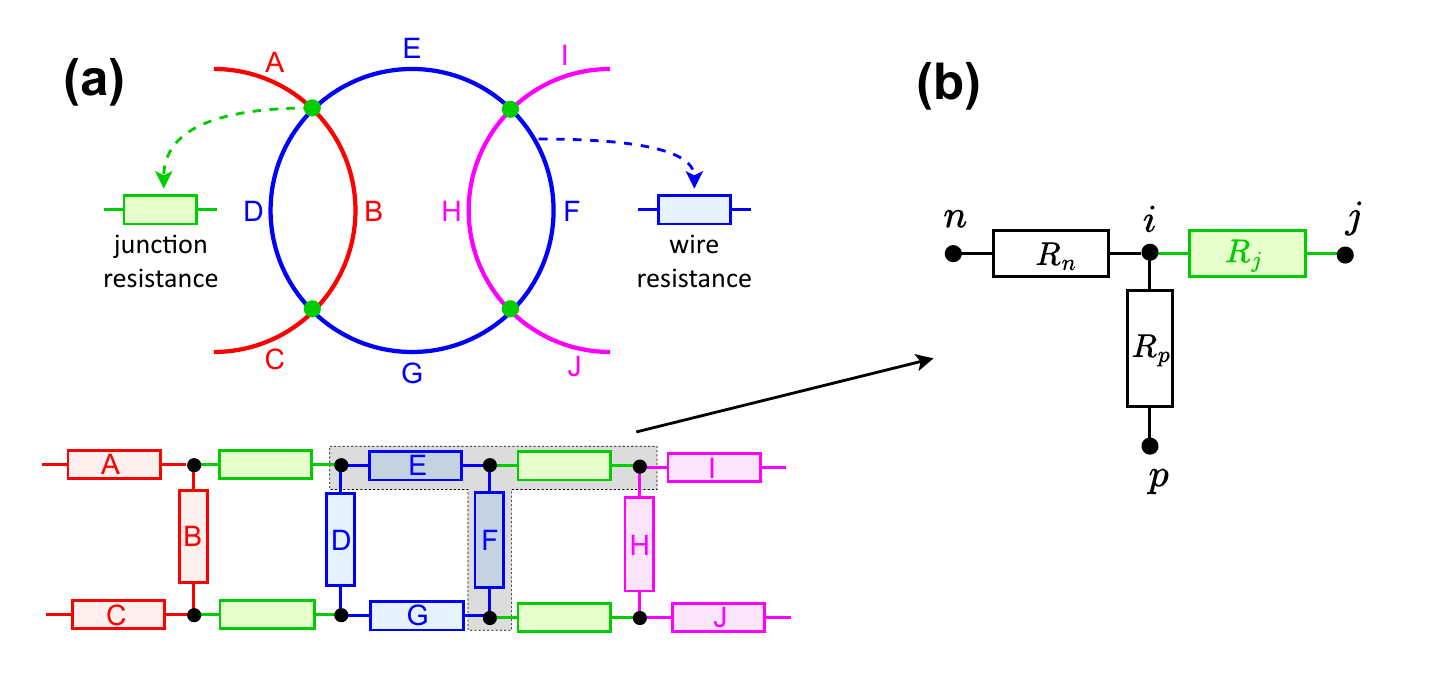}
    \caption{(a) A simple system of rings and their corresponding Equivalent Electrical Circuit (EEC). Intersecting rings create junctions (green dots). Every junction results in one junction resistance (green resistors) and 2 nodes (black dots) in the EEC. (b) Typical layout of a node $i$ in the network, surrounded by 3 nodes ($p$, $n$, $j$) and their respective segments ($R_p$, $R_n$, $R_j$)}
    \label{fig:EEC_and_KCL}
\end{figure*}

A voltage difference $V$ is applied between the busbars at the vertical boundaries of the sheet, and the horizontal boundaries are considered to be electrically insulating. To calculate the total resistance of the nanoring network, the Equivalent Electrical Circuit (EEC) is constructed for the spanning cluster. Only rings from the spanning cluster are considered, as rings from other clusters do not contribute to the conduction. A simple example of how an EEC is generated from a set of rings is shown in \cref{fig:EEC_and_KCL}(a). 

The EEC is constructed as follows. 
When two rings intersect, they share 2 junctions. 
At each junction, a node is constructed on both rings at the location of the junction. 
A ring then has a collection of nodes, and in general, each node in the ring is connected to its 2 neighboring nodes (on the same ring) with what will be referred to as wire segments. 
These segments represent the wires (arcs) of the ring with a wire resistance $R_w$ of
\begin{equation} 
    R_w = \rho \frac{l}{A}, \label{eq:wire_resistance}
\end{equation}
where $\rho$ is the wire resistivity, $l$ is the length of the wire (arc) and $A$ is the cross-sectional area of the wire. 
Note that even though the rings are assumed to have a zero thickness for calculating clusters (see \cref{ss:Geometrical_considerations}), $A$ is a non-zero input parameter of the simulation, and has an effect on the resistance of the wires according to \cref{eq:wire_resistance}. 
As Hamans et al. \cite{hamans_optical_2022} showed, when modeling the cross-section of nanowires, the exact shape of the cross-section has little to no impact on the $R_\square-T$ performance.
Therefore, a simple circular cross-section is chosen, and the rings have a wire thickness of $d=2\sqrt{A/\pi}$. 

Furthermore, each node on a ring is connected to a node on an overlapping ring that corresponds to the same junction, e.g., in \cref{fig:EEC_and_KCL}(a), the node on the red ring between wire segments A and B is connected to the node between wire segments D and E on the blue ring. 
Between these two nodes, a junction segment with a constant junction resistance $R_j$ represents the resistance of the contact between the rings.
The result of these operations is a graph that contains the structural information of the network, i.e., the EEC. 

The quantities of interest are the voltages at each node. 
To calculate them, Kirchhoff's Current Law (KCL) is applied to each node in the system.
To further illustrate this, \cref{fig:EEC_and_KCL}(b) shows a typical element of the network: a node $i$ is always connected to 3 other nodes by 3 segments.
One of these nodes ($j$, for ``junction") is a node on an intersecting ring, and it is connected to $i$ by a junction resistor $R_j$.
The other 2 nodes ($p$ and $n$, for ``previous" and ``next") are nodes from the same ring as node $i$, and they are connected to $i$ by wire resistors $R_p$ and $R_n$ respectively. 
Applying Kirchhoff's Current Law to node $i$ and then using Ohm's law, gives:

\begin{align}
    0 &= \frac{v_n - v_i}{R_n} + \frac{v_p - v_i}{R_p} + \frac{v_j - v_i}{R_j}\\
    &\Updownarrow\nonumber \\
     0 &= v_i\left( -\frac{1}{R_p}-\frac{1}{R_n}-\frac{1}{R_j}\right) \nonumber \\
     & \hspace{1cm}+ v_p\left(\frac{1}{R_p}\right) +  v_n\left(\frac{1}{R_n}\right) + v_j \left( \frac{1}{R_j}\right)\label{eq:KCL},
\end{align}
where $v_k$, is the voltage at the respective node $k\in\{i,n,p,j\}$. 

Repeating this process on each node in the system yields a set of linear equations, such as \cref{eq:KCL}. The set of equations is transformed into a matrix equation. Although this matrix can be very large (more than half a million rows for our simulations), it is also sparse, and it is solved with the SparseLU solver from the Eigen library, a \verb|C++| library for linear algebra \cite{guennebaud_eigen_2010}. The solution to this set of equations yields the voltage on every node, and the system is considered to be solved: the voltage, current, etc., can be calculated for every point in the system. An example of this result is shown in \cref{fig:gen_perc_cur}(c).
It is then possible to calculate the total current $I_{tot}$ flowing through the system, which is the sum of the currents between the vertical terminals. Combined with the given voltage difference $V$ between the left and the right vertical terminals, Ohm's law enables the calculation of the sheet resistance $R_{\square} = V/I_{tot}$.

Besides sheet resistance, the second macroscopic quantity of interest is the optical transparency $T$.
In literature, comparing transparency is often done by looking at transmittance values at a wavelength of 550 nm (green). 
The precise in silico calculation of $T$ is convoluted task, and therefore this work uses a simpler model to calculate $T$. 
The transparency of the system will largely depend on the Degree of Coverage (DoC) of the nanorings. 
This is the fraction of the sheet that is covered by the nanorings, whether they belong to the percolating cluster or not.
To estimate the optical transparency, $T = 1-$DoC is used.
Real-world systems will only approximately obey this relationship, but this simple model allows for a immediate, first-order investigation in the combined $R_\square-T$ performance of nanoring networks.

What is left then is to calculate the DoC. 
To do this, a classic Monte Carlo technique is used.
2D points are randomly sampled from a uniform distribution in the simulation domain $[0,L] \times [0, L]$.
For each point, it is checked if this point intersects with a ring (with thickness $d$) in the network. 
If this is the case, a variable $N_{hits}$ is increased by one. 
The fraction of the area of the sheet that is covered by the nanorings is then approximated by dividing $N_{hits}$ by the total amount of points that were sampled, $N_{sampled}$. That is,  DoC $\approx N_{hits}/N_{sampled}$.
By sampling more points, the fraction $N_{hits}/N_{sampled}$ converges to the exact value of the DoC. 
The Monte Carlo procedure stops when the difference between the smallest and highest values of the DoC, as based on the latest 10$^5$ sampled points, is less then $10^{-4}$. 
For each nanoring network, at least 10$^5$ points are sampled. 

The ideal transparent electrode maximizes both the conductance and the transparency of the network. 
However, in general, there is a trade-off between these two quantities for transparent electrodes. 
For example, using more rings (increasing $\eta$) will improve the conductance at a cost of a lower transparency, while using less rings will do the opposite. 
Therefore, this is a case of multi-objective optimization. 
One can combine the two quantities to define a Figure of Merit (FoM), which can be used to find an ``optimal configuration'' according to the definition of the used FoM. For example, a FoM that is often used in literature \cite{azani_transparent_2019, yun_fabrication_2013, vasudevan_fabrication_2026, kim_flash-lamp_2026, yu_flexible_2026, lagrange_optimization_2015, zhu_flexible_2019} is the one defined by Haacke \cite{haacke_new_1976}, who originally defined it for transparent conductive thin films:
\begin{equation} \label{eq:FoM}
    \text{FoM} = \frac{T^{10}}{R_\square}.
\end{equation}
It is important to note here that the exact expression of the FoM, as well as the exponent of $T$ in particular, is quite arbitrary.
The exact expression of the FoM is application-dependent, as certain applications require a larger emphasis on transparency instead of conductance, while for other applications this is the other way around. 
As an example, if the transmittance of a material in a certain application is more crucial than \cref{eq:FoM} reflects, one can increase the exponent of $T$ to e.g. 20 or even 100. 
The network which maximizes the resulting FoM will then have a higher $T$ compared to the network that maximizes the FoM of \cref{eq:FoM}.

A FoM can be used to assess the $R_\square-T$ performance of a transparent electrode with a single number, however it does not tell the whole story. 
One still has to take into account the minimal requirements of the desired application.
Keeping this in mind, we include the Haacke FoM of \cref{eq:FoM} in the performance assessment of the simulated networks in \cref{s:results} as an example FoM.

\subsection{Electrical damage} \label{ss:electrical_damage}

Nanoring networks are susceptible to breakdown when in use due to e.g. localized Joule heating  \cite{khaligh_failure_2013}. 
To model this behavior, a procedure comparable to that of Charvin et al. \cite{charvin_dynamic_2021} is utilized, as it provides a straightforward yet effective reproduction of experimental results. 
The process of applying electrical damage starts with the undamaged system. 
After calculating the voltages, currents, etc. as described in \cref{ss:R-T_performance}, the segment (either a wire segment or a junction segment) in the network with the highest current is removed from the system, as it is a segment that is likely to get destroyed by electrical damage.
Then, the voltages and currents in the damaged system are recalculated and the process repeats.
After a certain amount of segments have been removed, the system will no longer contain a percolating cluster, and the electrode is destroyed.

Another interesting way to model this electrical breakdown is to systematically remove the segment with the highest power instead of the one with the highest current. This is because the Joule effect is directly proportional to the power. However, previous works that use the model only ever use the current-based destruction criterion. This work performs both types of degradation simulations (current-based and power-based) separately. To keep the discussion concise, the breakdown based on electrical current is discussed in the main text (\cref{ss:electrical_damage}) while the power-based breakdown (and its differences with the current-based breakdown results) is discussed in the Supplementary Information.

\section{Results and Discussion} \label{s:results}
\FloatBarrier
\subsection{Parametric investigation} 
\begin{figure*}
    \centering
    \includegraphics[height=0.9\textheight, keepaspectratio]{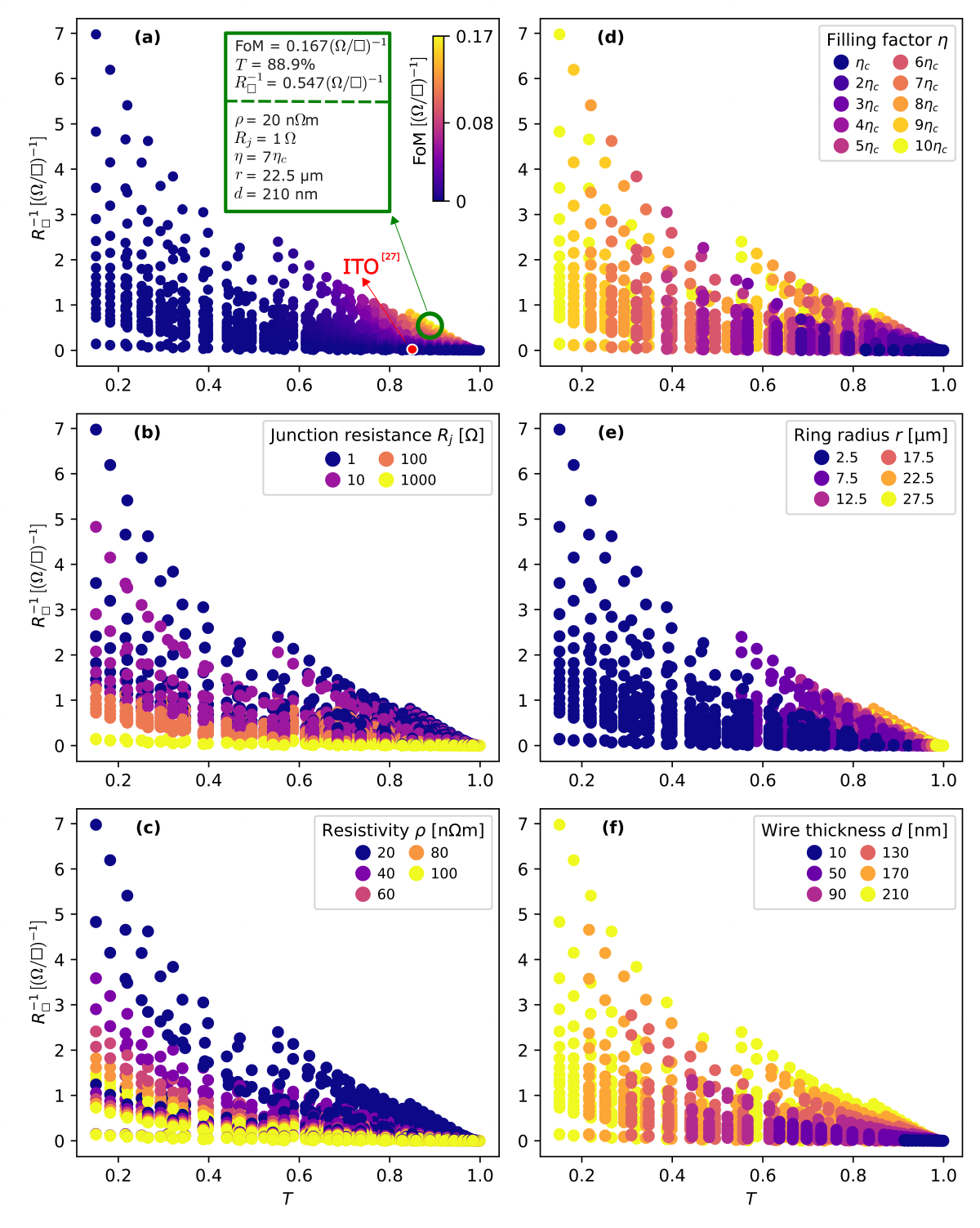}
    \caption{Overview of the parametric investigation. For each combination of the parameter grid shown in \cref{tab:pareto_front_overview}, the average sheet conductance $R_\square^{-1}$ and transparency $T$ of 1000 systems is shown as a point on each of the six subplots (a)-(f). Standard deviation error bars are present in both vertical and horizontal directions, but are smaller than marker size. In (a), the points are colored by Figure of Merit using \cref{eq:FoM}. The green circle marks the point with the highest FoM. The red dot marks the performance of ITO \cite{mazur_influence_2010}.  In (b)-(f), the points are colored by the five parameters that are varied.}
    \label{fig:pareto_front_overview}
\end{figure*}

In order to examine the effect of multiple design parameters on the performance indicators ($R_\square$ and $T$), a parameter grid search is performed on square systems with size $L=100$ \textmugreek m. 
The parameter grid can be found in \cref{tab:pareto_front_overview}.
The ranges of these parameter values are chosen based on published data of metallic nanorings and nanowires, summarized in Table I and Table II in Tarasevich, 2023 \cite{tarasevich_transparent_2023}. 
For each parameter combination, the average sheet conductance $R_\square^{-1}$ and transparency $T$ of 1000 networks is calculated and displayed as a dot in every subplot of \cref{fig:pareto_front_overview}. 

In \cref{fig:pareto_front_overview}(a), the performance of the networks are colored by the Figure of Merit (FoM) in \cref{eq:FoM}. 
This FoM places a high emphasis on transparency: the parameter combinations that result in  $T<80\%$ have an insignificant FoM. 
In the bottom right of the plot, there are still systems that display a high level of transparency ($T>99\%$), but their sheet resistance is too high and therefore their FoM is close to zero.
The system with the highest FoM is marked with a green circle in the figure. 
The figure shows that there are a lot of nanoring systems that exhibit a better performance than that of ITO. 

In the 5 other subplots (b)-(f), the same systems are plotted as in subplot (a), but the color now reveals the parameters that were used. 
For example, take the system on the top left of the plots (where $R_\square^{-1} \approx 7\;(\Omega/\square)^{-1}$. 
Subplot (a) shows that this system results in a low FoM, as it is colored in a dark blue. 
To find the values of $R_j$, $\rho$, $\eta$, $r$ and $d$ that were used in this network, we identify the same point in the 5 subplots (b)-(f), and look at the corresponding color legend.
Therefore it is revealed that this network used $R_j=1$ $\Omega, \rho = 20$ n$\Omega$m, $\eta=\;10\eta_c, r=$ 2.5 \textmugreek m, $d=$ 210 nm. 

Inspecting \cref{fig:pareto_front_overview}(b), it is shown that increasing $R_j$ has the evident effect of increasing the sheet resistance, without affecting the transparency. 
For example, increasing $R_j$ from 1 $\Omega$ (blue points) to 10 $\Omega$ (purple points), simply shifts the points down on the figure. 
The same effect is found for the resistivity $\rho$ in \cref{fig:pareto_front_overview}(c). 

Changing the filling factor $\eta$ is more interesting, as this has an effect on both the sheet resistance and the transparency.
Increasing the filling factor shifts the data points towards the top left of the plot, as more rings in the system means more rings are generated, improving the resistance at the cost of transparency. 
The ideal filling factor therefore depends on the specific requirements of the application. 
The maximal value of the FoM used in \cref{eq:FoM} was reached for a filling factor equal to $7\eta_c$.

Decreasing the ring radius $r$ at the same value of the filling factor \cref{eq:filling_factor} results in more rings being generated in the system. Therefore, increasing the ring radius $r$ has the opposite effect, increasing transparency at the cost of sheet resistance. 
However, this increase in transparency slows down for larger radii, and for the largest radius, the transparency no longer improves. The FoM was optimal for the second largest radius, $r=22.5$ \textmugreek m. 

As expected, increasing the wire thickness $d$  qualitatively has the same effect as increasing the filling factor, moving the points to the upper left of the plot. 
At the lowest values of the wire thickness, the transparency is maximized, however the sheet resistance is too high for most relevant applications. 
Additionally, as Lagrange et al. \cite{lagrange_optimization_2015} showed, thinner wires lead to a faster degradation due to electrical damage, so one has to keep this in mind. 
For the optimal FoM, the wire thickness was largest: $d = 210$ nm. If one were to increase the thickness even more, it is plausible to find larger FoM values. However, the parameter ranges in \cref{tab:pareto_front_overview} is chosen in accordance with the literature, taking into account available experimental data \cite{tarasevich_transparent_2023}. 
\break

\begin{table}
\centering
 \caption{Parameter grid used to create \cref{fig:pareto_front_overview}. (*) The step size of the junction resistance was multiplicative instead of additive.}
 \label{tab:pareto_front_overview}
  \begin{tabular}{@{}llll@{}}
    \hline
    Parameter name & First value & Last value & Step size \\
    \hline
    Filling factor $(\eta)$  & $\eta_c$  & 10 $\eta_c$  & $\eta_c$\\
    Ring radius $(r)$ & 2.5 \textmugreek m & 27.5 \textmugreek m & 5 \textmugreek m  \\
    Wire thickness $(d)$ & 10 nm & 210 nm & 40 nm \\
    Junction resistance $(R_j)$ & 1 $\Omega$ & 1000 $\Omega$ & 10 (*) \\
    Wire resistivity ($\rho$) & 20 n$\Omega$m & 100 n$\Omega$m & 20 n$\Omega$m\\
    \hline
  \end{tabular}
\end{table}

\begin{figure*}
    \centering
    \includegraphics[width=1\linewidth]{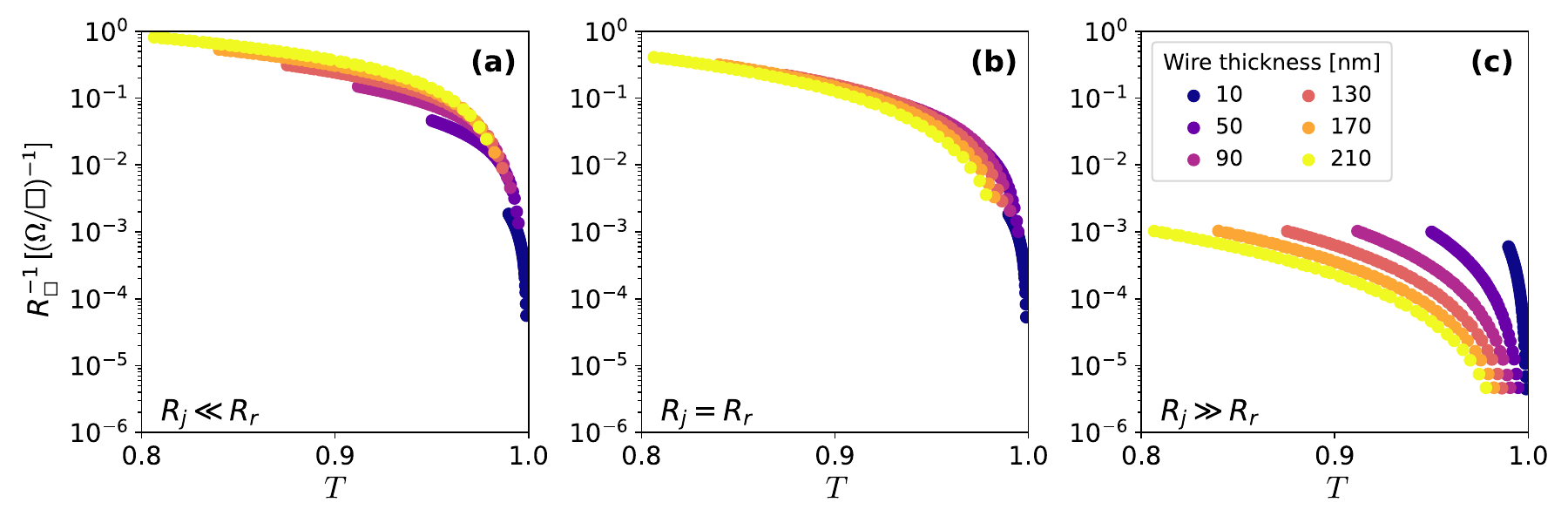} 
    \caption{$R_\square-T$ performance of networks for different values of the wire thickness (see legend). For each wire thickness, varying values of the filling factor are used, ranging from $\eta_c$ to $10\eta_c$ with a delta of $0.2\eta_c$. Each data point is the average result of 1000 simulations. Simulations are performed in three resistance regimes: (a) wire dominated, (b) mixed and (c) junction dominated.}
    \label{fig:A_and_Rj}
\end{figure*}

\Cref{fig:A_and_Rj} shows the impact of the wire thickness for networks of size $L=100$ \textmugreek m with varying filling factors in three resistance regimes (wire dominated, mixed and junction dominated regime).
To define these resistance regimes, a characteristic resistance value $R_r$ of a single nanoring is obtained by taking a typical silver nanoring ($r=20$ \textmugreek m, $\rho=22.6$ n$\Omega$m, $d=100$ nm) and picturing that the opposite sides of the ring are connected to the two ends of a voltage source. 
The ring is thus split up in two equal halves, and the equivalent resistance of the ring $R_r$ is the combined resistance of the two (parallel) half rings.
Following \cref{eq:wire_resistance}, each of these halve rings has a wire resistance equal to $\pi r\rho/A$ where $A=\pi\left(d/2\right)^2$ is the wire cross-sectional area.

As these are parallel resistors, the characteristic resistance of a single nanoring is
\begin{equation}
    R_r = \frac{1}{\left(\pi r\rho/A\right)^{-1} + \left(\pi r\rho/A\right)^{-1}} =\frac{\pi r \rho}{2A} = \frac{2 r\rho}{d^2}
\end{equation}
The wire dominated, mixed and junction dominated resistance regimes are then defined in this work as the regimes where $R_j = R_r\times10^{-3}$, $R_j = R_r$ and $R_j = R_r\times10^{3}$ respectively.

For these simulations, the filling factor varies from $\eta_c$ to $10\eta_c$ with steps of $0.2 \eta_c$. The ring radius and wire resistivity were kept constant at 20 \textmugreek m and 2.26$\times 10^{-8} \Omega$m respectively. Sheet resistance and transparency values are averaged over 1000 realizations.
In the first two resistance regimes, increasing the wire thickness has the same effect, improving the resistance at the cost of transparency.
This is due to the resistance of wire segments being inversely proportional to the cross-section of the wire, therefore decreasing the wire resistance and thus positively impacting the sheet resistance.
On the contrary, increasing the wire thickness decreases the transparency as more material is deposited on the sheet, increasing the DoC. 
In the junction dominated case however, increasing the wire thickness only decreases the transparency without any significant improvement in the resistance. 
In this case, the sheet resistance is mostly due to the junctions, while the wire segments barely contribute. 
Thus, increasing the wire thickness will decrease the wire resistances, but the macroscopic effect on $R_\square$ is negligible. 
Although not affecting the sheet resistance, increasing the wire thickness still negatively affects the transparency.
Therefore, in the case of badly optimized junctions, the advice is to create thin nanorings.

\subsection{Breakdown and crack formation}

\begin{figure*}
    \centering
    \includegraphics[width=\linewidth]{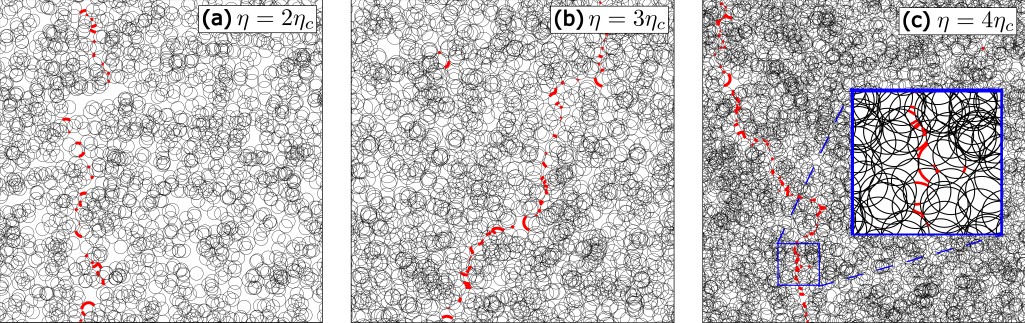}
    \caption{Examples of networks that have been damaged by the breakdown algorithm with varying filling factors. The segments that carried the highest current (and thus were destroyed in the algorithm) are marked in red. Values of filling factors are $2\eta_c$, $3\eta_c$ and $4\eta_c$ for (a), (b) and (c) respectively. }
    \label{fig:cracks}
\end{figure*}

Using the procedure described in \cref{ss:electrical_damage}, networks are repeatedly damaged until electrode failure, meaning no percolating cluster is present.
\Cref{fig:cracks} illustrates the formation of cracks in the network as a result of this breakdown process. 

For these simulations, parameter values $L=100$ µm, $r=2.5$ µm, $d=100$ µm, $R_j=30\;\Omega$ and $\rho = 22.6$ n$\Omega$m are used, and three different filling factors are used ($2\eta_c, 3\eta_c$ and $4\eta_c$). 

In all cases, the formation of a quasi-vertical crack is found: the segments that were destroyed during the breakdown simulations seem to follow a line that is parallel to the busbars. 
The other segments (not in the vicinity of the crack) are largely unaffected: the damage starts localized and propagates throughout the whole system, in the general direction parallel to the busbars. 
This is characteristic of electrical damage and it is contrary to the global destruction that is representative of the thermal damages that arise from high-temperature thermal annealing \cite{langley_metallic_2014}. 

The appearance of the vertical cracks are consistent with previously reported experimental and simulated degradation studies for nanowire networks, supporting the validity of the used degradation model \cite{charvin_dynamic_2021, zhu_improving_2020, grazioli_predicting_2024}. 
For the first time, this work confirms that the crack formation due to electrical damage is also present in nanoring electrodes, similar to those in nanowire electrodes. 

The formation of the vertical crack is explained by the fact that each vertical cross-sectional slice of the network has to carry the same amount of current (as a consequence of charge conservation).  
When a previous segment has been destroyed in the network, the other segments in the vertical slice must also carry a current that the broken segment no longer can. Therefore, the segments in this vertical slice have to carry a disproportional larger amount of current compared to segments in other vertical slices. This means that the segment with the highest current in the next iteration of the breakdown algorithm will likely be in the same vertical slice as the previously destroyed segment, creating a vertical crack. 

During the breakdown process, the simulation records the sheet resistance. 
To display the degradation of the conductance of the network, the electrical integrity $I_e$ is defined as the ratio between the sheet resistance in the undamaged state and its resistance in the damaged state, comparable to how Grazioli defined it for nanowire networks \cite{grazioli_predicting_2024}. 
$I_e$ is a dimensionless quantity that takes on values between 1 and 0, where $I_e=1$ represents the undamaged, intact system. During the electrical breakdown, the sheet resistance will increase as a result of the removal of conducting segments. This will result in a decrease of $I_e$ until a conducting pathway no longer exists, meaning $R_\square$ diverges and thus $I_e=0$.

\begin{figure*}
    \centering
    \includegraphics[width=1\linewidth]{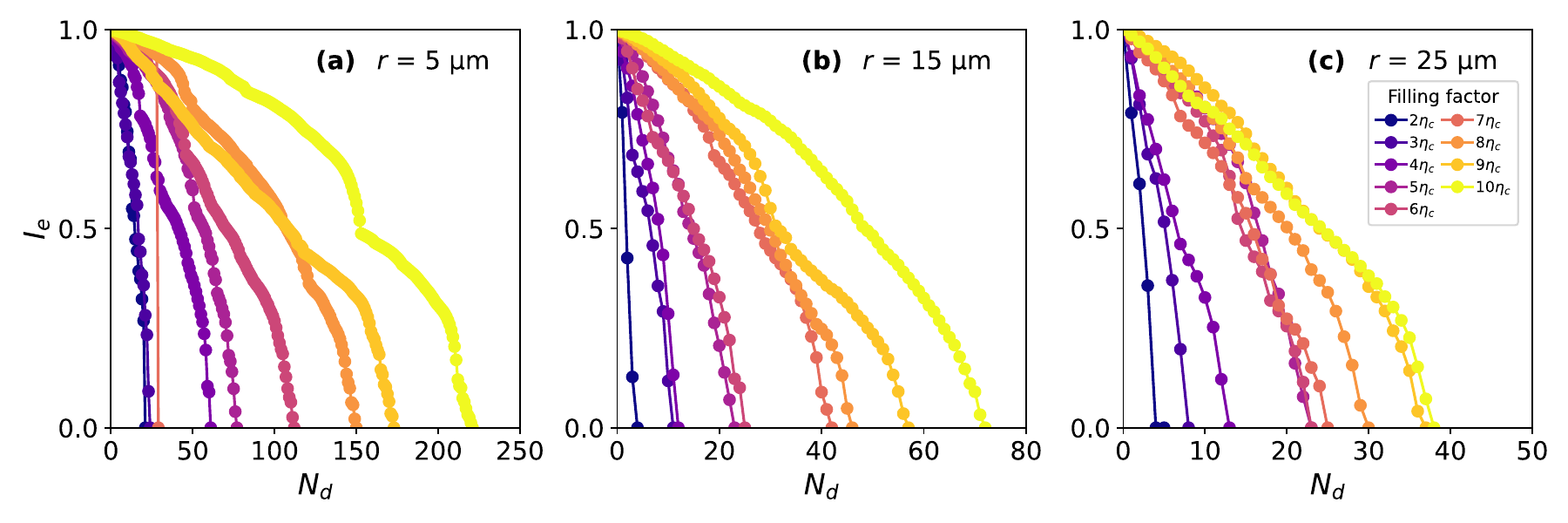}
    \caption{Electrical integrity $I_e$ during the breakdown simulation in function of the number of damaged segments $N_d$. In each plot, the collection of points of one color represent the electrical integrity of a single nanoring network. In each plot, $I_e$ is shown for different filling factors, ranging from $2\eta_c$ (blue) to $10\eta_c$ (yellow). The three plots display $I_e$ for three different values of the ring radius $r$. Note the different horizontal axis limits. }
    \label{fig:e_dmg_normal}
\end{figure*}

\Cref{fig:e_dmg_normal} shows the degradation of a system in function of the number of destroyed segments $N_d$ for different values of the filling factor $\eta$  and the ring radius $r$. Filling factors $\eta$ from $2\eta_c$ to $10\eta_c$ are used in steps of $\eta_c$ and radii $r=$ 5 \textmugreek m, 15 \textmugreek m and 25 \textmugreek m are chosen. Other parameters were kept constant at $L=100 $ \textmugreek m, $R_j=30\;\Omega, \rho = 22.6$ n$\Omega$m and $d=$ 200 nm. 

Clearly, higher filling factors result in more electrically stable networks, as $I_e$ varies more slowly with $N_d$.
As a result of this, the amount of segments needed to completely destroy the system $N_d^*$ also increases.

When comparing systems between different ring radii, $r$ affects the decay of $I_e$ substantially (note the horizontal axis limits in \cref{fig:e_dmg_normal}), and an inverse relationship between $r$ and $N_d^*$ is identified.
The networks that are most resistant to the electrical damage are dense systems with small rings.

The shape of the $I_e$ curves is quite remarkable. 
Firstly, the curves are all quite similar to each other (ignoring some statistical noise and a few sudden drops). 
$I_e$ generally displays a linear decrease until a certain threshold value of around $I_e = 0.3$, where all curves show a sudden sharp drop-off. 
This effect has also been reported both experimentally and in simulations of nanowire networks \cite{khaligh_failure_2013, charvin_dynamic_2021, grazioli_predicting_2024, resende_time_2022}.
This work shows that the effect is not exclusive to nanowire networks, but is present in nanoring networks as well.

\begin{figure*}
    \centering
    \includegraphics[width=1\linewidth]{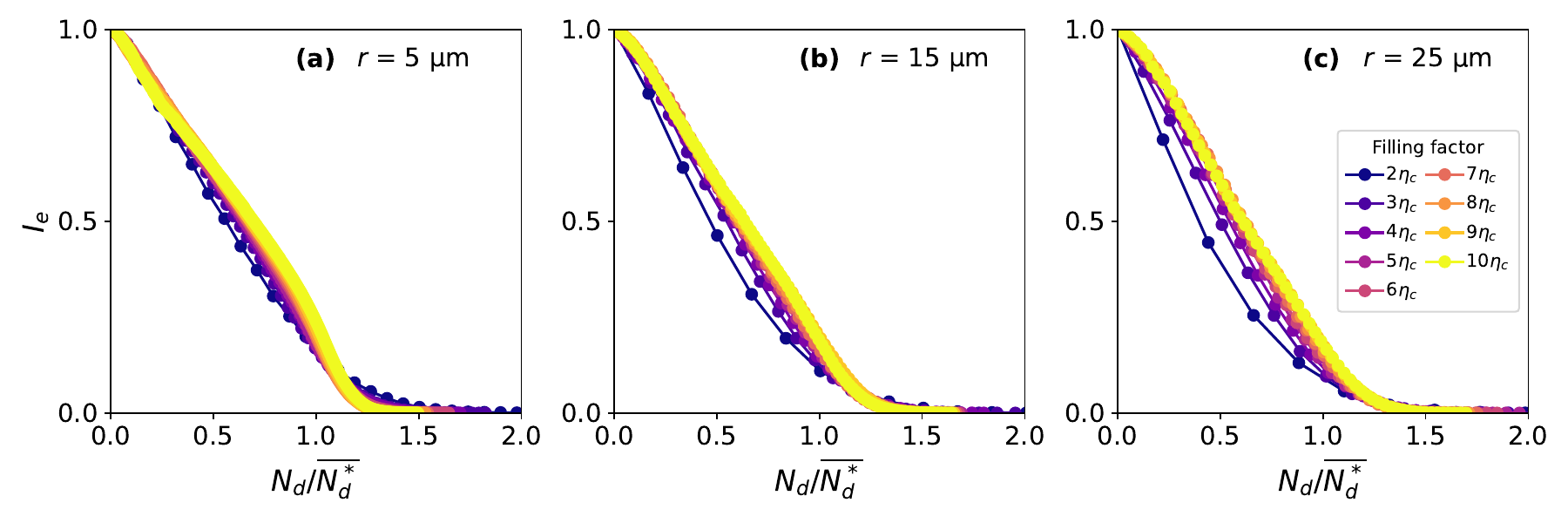}
    \caption{Electrical integrity $I_e$ during the breakdown simulation in function of the number of damaged segments $N_d$, scaled with the average number of segments needed to destroy the network $\overline{N^*_{d}}$. In each plot, the collection of points of one color represents the average electrical integrity of 1000 networks. In each plot, $I_e$ is shown for different filling factors, ranging from $2\eta_c$ to $10\eta_c$. The three plots display $I_e$ for three different values of the ring radius $r$.}
    \label{fig:e_dmg_scaled}
\end{figure*}

One can note that the degradation for different filling factors looks quite similar.
The different graphs appear to differ only in a horizontal scaling.
To investigate this, a series of breakdown simulations are performed (using the same parameter values as described earlier in this section), and the values of the electrical integrity are averaged over 1000 runs. 
Then, for each value of the filling factor, the horizontal axis is scaled by division of the average number of segments needed to completely destroy the network $\overline{N^*_{d}}$.
The resulting plots in \cref{fig:e_dmg_scaled} show that using this scaling, the $I_e$ curves are identical to within a substantial degree of precision.
Therefore, the relative electrical breakdown of nanoring networks, as measured by its sheet resistance, is universal with respect to the filling factor. 
This universality occurs regardless of ring radius value.

\section{Conclusion} \label{s:conclusion}
 Metallic nanocomposite networks such as those made from nanowires and nanorings are promising flexible, transparent conductive materials for emerging technologies. In this work we focus on metallic nanoring networks and (i) their balance between electrical conductance and optical transparency, as well as (ii) their breakdown due to electrical damages when in use. This work used an in-house developed \verb|C++| code to simulate a large amount of conductive nanoring networks. Both the sheet resistance and the transparency are modeled: the former by transforming the network into an Equivalent Electrical Circuit (EEC) where both the resistance of the ring and its contacts are taken into account, and the latter by using a Monte Carlo method to estimate the degree of coverage.
 
In the first part of the work, a parametric study is performed where 5 parameters (both geometric and material properties) were varied, and its effect on the sheet resistance and transparency are investigated using the Haacke Figure of Merit. For our cases, it is found that the network with optimal FoM is the one with a filling factor $\eta = 7\eta_c$, a ring radius $r=22.5$ \textmugreek m and a wire thickness $d=$ 210 nm. Additionally, it is also shown that in the junction dominated resistance regime, although the wire thickness affects the transparency, it does not significantly alter the sheet resistance.

In the second part of the work, the breakdown due to electrical damage of the nanoring networks are modeled for the first time. The networks display cracks parallel to the busbars at the vertical sheet boundaries. Additionally, the degradation of the sheet resistance is measured. Higher density networks with smaller rings required the largest number of segments to be damaged in order for the network to lose its conductive behavior. The relative degradation of the sheet resistance is shown to be universal for networks with different filling factors, and this is observed for all values of the ring radius.

\medskip
\noindent\textbf{Acknowledgments}\\
This study was supported by the Special Research Fund (BOF) of Hasselt University using BOF number BOF24OWB29. The resources and services used in this work were provided by the VSC (Flemish Supercomputer Center), funded by the Research Foundation - Flanders (FWO) and the Flemish Government.

\medskip
\noindent\textbf{Conflict of Interest}\\
The authors declare no conﬂict of interest.

\medskip
\noindent\textbf{Data Availability Statement}\\
Data (simulation output) used to create the plots in this paper, as well as the simulation code itself can be found using the following DOI: 10.5281/zenodo.20125324. Additionally, the data and simulation code are accompanied by an interactive notebook (.ipynb) that creates the plots from the simulation output.

\bibliography{references}

\appendix*
\section{Modeling electrical damage of metallic nanoring networks by inspecting segment-specific power values}

As stated in \cref{ss:electrical_damage} in the main text, one can model the electrical breakdown of nanoring networks by solving the set of linear equations resulting from Kirchhoff's Current Law (KCL), removing the segment (either a wire segment or a junction segment) in the system that has the highest electrical current and repeating this process untill complete electrode failure. 
Joule heating is however directly proportional to the power $P = UI$, where $U$ is the voltage difference between both ends of the segment and $I$ is the current flowing through it. 
Therefore the breakdown simulations are repeated here, where now the segment with the highest power is removed (instead of the segment with the highest current). 
With the exception of this modification in the breakdown protocol, the utilized methods and parameter values of the power-based breakdown simulations are identical to those of the current-based ones described in the main text. 

\begin{figure*}
    \centering
    \includegraphics[width=0.9\linewidth]{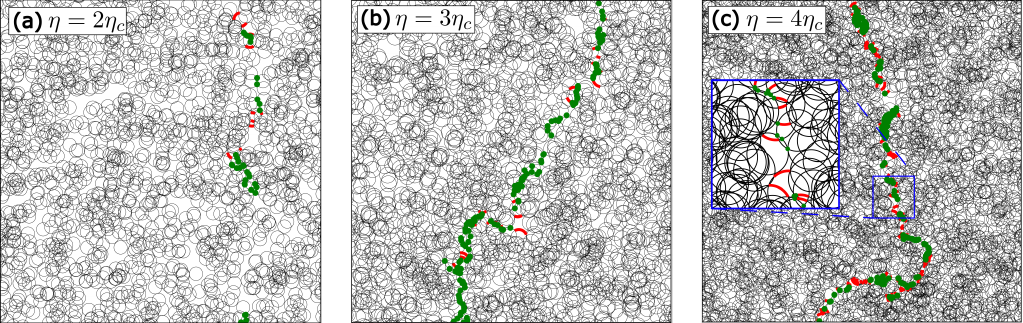}
    \caption{Examples of networks that have been damaged by the breakdown algorithm with varying filling factors. The wire segments and junction segments that were destroyed in the algorithm are marked in red lines and green dots respectively. Values of filling factors are $2\eta_c$, $3\eta_c$ and $4\eta_c$ for (a), (b) and (c) respectively.}
    \label{fig:cracks_power}
\end{figure*}

\begin{figure*}
    \centering
    \includegraphics[width=0.9\linewidth]{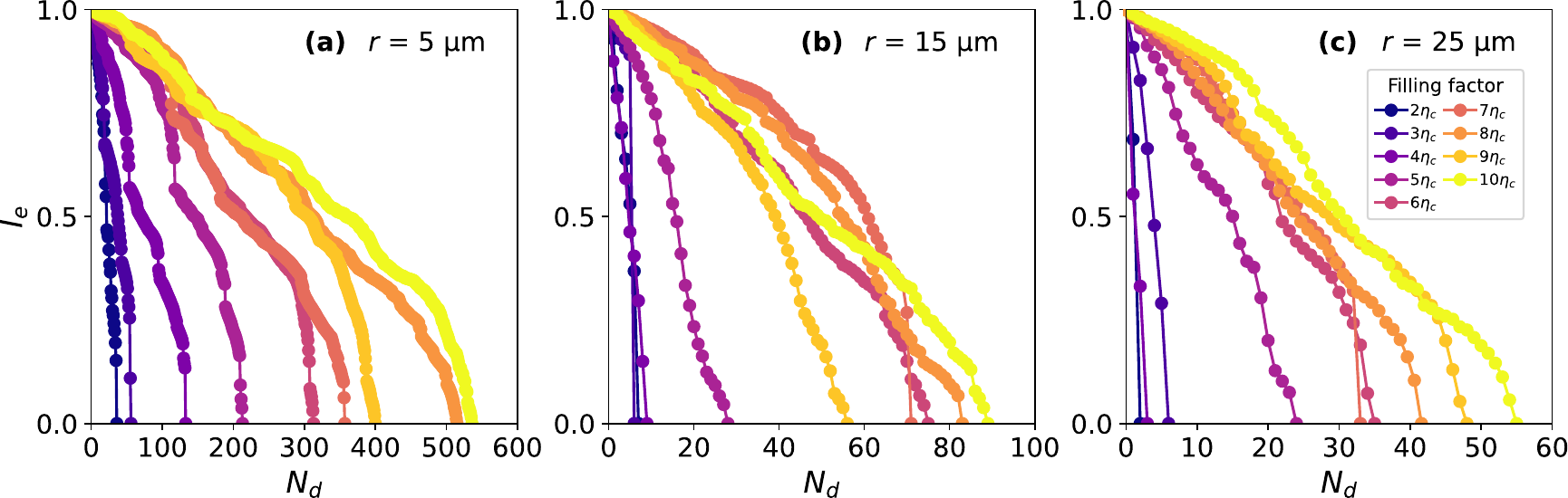}
    \caption{Electrical integrity $I_e$ during the breakdown simulation in function of the number of damaged segments $N_d$. In each plot, the collection of points of one color represent the electrical integrity of a single nanoring network. In each plot, $I_e$ is shown for different filling factors, ranging from $2\eta_c$ (blue) to $10\eta_c$ (yellow). The three plots display $I_e$ for three different values of the ring radius $r$. Note the different horizontal axis limits. }
    \label{fig: e_dmg_normal_power}
\end{figure*}

\begin{figure*}
    \centering
    \includegraphics[width=0.9\linewidth]{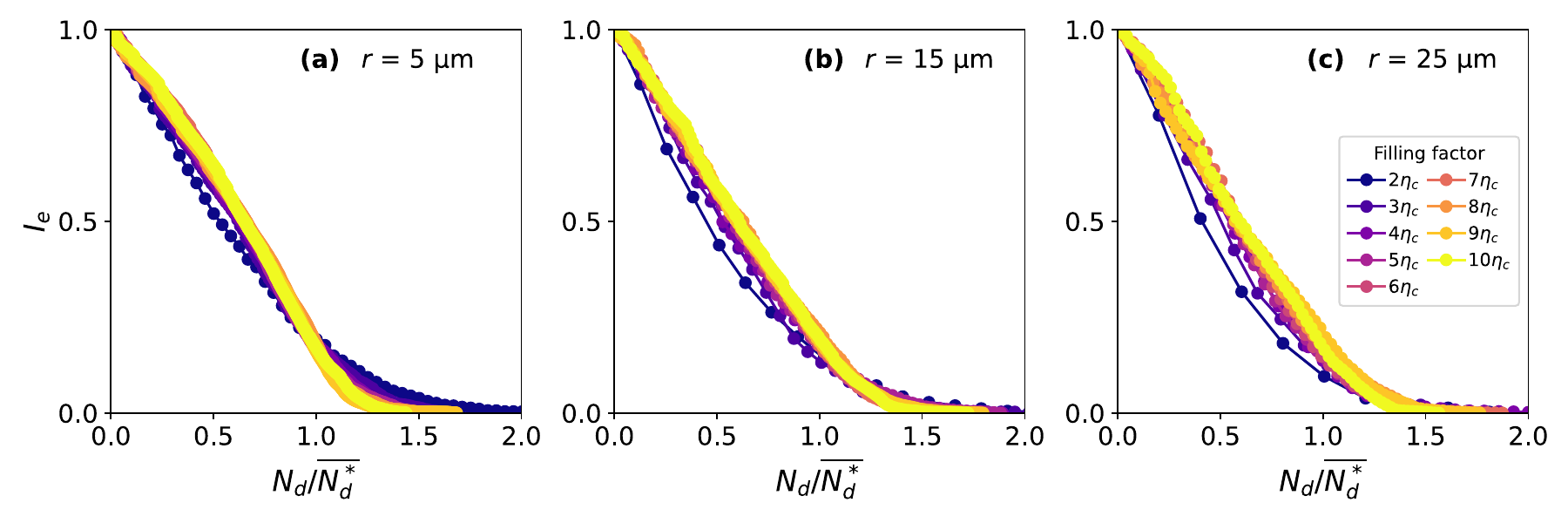}
    \caption{Electrical integrity $I_e$ during the breakdown simulation in function of the number of damaged segments $N_d$, scaled with the average number of segments needed to destroy the network $\overline{N^*_{d}}$. In each plot, the collection of points of one color represents the average electrical integrity of 100 networks. In each plot, $I_e$ is shown for different filling factors, ranging from $2\eta_c$ to $10\eta_c$. The three plots display $I_e$ for three different values of the ring radius $r$.}
    \label{fig: e_dmg_scaled_power}
\end{figure*}

\subsection{Crack formation}
\Cref{fig:cracks_power} shows three examples of networks after they have been completely destroyed by the electrical damage simulations, and thus are no longer able to conduct. 
As is the case in the breakdown simulations based on the removal of the highest current segment, a quasi-vertical crack is observed. 
If the destroyed segments were to be erased from the figure, one can travel from the bottom to the top of the electrode without ever touching a nanoring, and thus the system is no longer percolating. 

In the power-based breakdown, a considerable fraction of the destroyed segments are junction segments. 
This is different from the current-based breakdown simulations, where almost exclusively wire segments are destroyed. 

\subsection{Electrical integrity}
In \cref{fig: e_dmg_normal_power}, the electrical integrity $I_e$ is shown throughout the breakdown simulation for individual nanoring networks with different values of the filling factor $\eta$ and ring radius $r$. 
Some similarities can be found between these $I_e$-profiles and the ones from the current-based breakdown simulations:
\begin{enumerate} 
    \item Although quite noisy, the graphs indicate that using a higher filling factor will increase the required amount of damaged segments to reach electrode failure $N_d^*$. 
    \item In the bulk of the $I_e$-profile, the integrity generally drops linearly with $N_d$. 
    \item Near the end of the simulation, $I_e$ decreases significantly, revealing a sharp increase in sheet resistance.
    \item An inverse relationship is found between $r$ and $N_d^*$.
\end{enumerate}

These $I_e$-profiles decrease more slowly with $N_d$ than the profiles from the current-based breakdown simulations. 
The amount of segments needed to stop the network from percolating $N_d^*$ is higher in the power-based simulations, and significantly higher for the smallest ring radius. 
This could be attributed to the fact that in the power-based simulations, a significant amount of the destroyed segments are of the junction type. 
Junction segments are point-like, while wire segments correspond to finite arcs on a ring.
Therefore, the spatial extent of wire segments is higher, and their destruction leads to a more pronounced increase in the size of the crack. 
As the current-based simulations almost exclusively destroy wire segments, less segments are needed to stop percolation, and thus $N_d^*$ is lower than in the power-based simulations.

\Cref{fig: e_dmg_scaled_power} shows the average electrical integrity $I_e$ of 100 simulations where the curves with different filling factors are horizontally scaled with the average number of segments needed to destroy the network $\overline{N^*_{d}}$ for that filling factor. 
Just like in the current-based breakdown simulations, the procedure is applied to systems with a range of filling factors ($2\eta_c$ to $10\eta_c$) and ring radii (5, 15 and 25 \textmugreek m). 
 Comparable to the current-based breakdown, the $I_e$ profiles are identical to a substantial degree. 
 Additionally, this behavior is present for all three values of the ring radius $r$.

\end{document}